\documentclass[longauth]{aa}                                                                                        
\usepackage{graphicx}                                                                                              
\usepackage{epsfig}                                                                                                
\usepackage{natbib}                                                                                                
\usepackage{lscape}                                                                                                
\usepackage{rotating}                                                                                              
\usepackage{tabularx}                                                                                              
\usepackage{amsmath}                                                                                              
\usepackage{subfig}
\usepackage{txfonts}
\usepackage{color}      
\bibpunct{(}{)}{;}{a}{}{,}

\def\mpcoh{\,h^{-1}{\rm Mpc}}
\def\kms{\,{\rm km\, s ^{-1}}}

\begin{document}
\title{The VIMOS Public Extragalactic Redshift Survey
\thanks{Based on
observations collected at the European Southern Observatory, Cerro Paranal, Chile, using the Very Large Telescope under programs 182.A-0886 and partly 070.A-9007.
Also based on observations obtained with MegaPrime/MegaCam, a joint project of CFHT and CEA/DAPNIA, at the Canada-France-Hawaii Telescope (CFHT), which is operated by the
National Research Council (NRC) of Canada, the Institut National des Sciences de l'Univers of the Centre National de la Recherche Scientifique (CNRS) of France, and the University of Hawaii. This work is based in part on data products produced at TERAPIX and the Canadian Astronomy Data Centre as part of the Canada-France-Hawaii Telescope Legacy Survey, a collaborative project of NRC and CNRS.}}
\subtitle{Measuring the growth rate of structure around cosmic voids}

\titlerunning{Measuring the growth rate of structure around cosmic voids}

\author{
A. ~J.~Hawken\inst{1}
\and B.~R.~Granett\inst{1}
\and A.~Iovino\inst{1}
\and L.~Guzzo\inst{1,2}         
\and J.~A.~Peacock\inst{14}
\and S.~de la Torre\inst{4}     
\and B.~Garilli\inst{3}     
\and M.~Bolzonella\inst{9}           
\and M.~Scodeggio\inst{3} 
\and U.~Abbas\inst{5}
\and C.~Adami\inst{4}
\and D.~Bottini\inst{3}
\and A.~Cappi\inst{9,21}
\and O.~Cucciati\inst{17,9}           
\and I.~Davidzon\inst{4,9} 
\and A.~Fritz\inst{3}
\and P.~Franzetti\inst{3}
\and J.~Krywult\inst{15} 
\and V.~Le Brun\inst{4}
\and O.~Le F\`evre\inst{4}
\and D.~Maccagni\inst{3}
\and K.~Ma{\l}ek\inst{16,23}
\and F.~Marulli\inst{17,18,9}
\and M.~Polletta\inst{3}
\and A.~Pollo\inst{22,23}
\and L.~A.~.M.~Tasca\inst{4}
\and R.~Tojeiro\inst{11}
\and D.~Vergani\inst{25,9}
\and A.~Zanichelli\inst{26}
\and S.~Arnouts\inst{6} 
\and J.~Bel\inst{7}
\and E.~Branchini\inst{10,28,29}
\and G.~De Lucia\inst{13}
\and O.~Ilbert\inst{4}
\and L.~Moscardini\inst{17,18,9}
\and W.~J.~Percival\inst{11}
}
\institute{
INAF - Osservatorio Astronomico di Brera, Via Brera 28, 20122 Milano, via E. Bianchi 46, 23807 Merate, Italy %1
\and  Universit\`{a} degli Studi di Milano, via G. Celoria 16, 20130 Milano, Italy %2
\and INAF - Istituto di Astrofisica Spaziale e Fisica Cosmica Milano, via Bassini 15, 20133 Milano, Italy%3
\and Aix Marseille Universit\'e, CNRS, LAM (Laboratoire d'Astrophysique de Marseille) UMR 7326, 13388, Marseille, France  %4
\and INAF - Osservatorio Astronomico di Torino, 10025 Pino Torinese, Italy %5
\and Canada-France-Hawaii Telescope, 65--1238 Mamalahoa Highway, Kamuela, HI 96743, USA %6
\and Aix Marseille Universit\'e, CNRS, CPT, UMR 7332, 13288 Marseille, France   %7
\and Universit\'{e} de Lyon, F-69003 Lyon, France %8
\and INAF - Osservatorio Astronomico di Bologna, via Ranzani 1, I-40127, Bologna, Italy %9
\and Dipartimento di Matematica e Fisica, Universit\`{a} degli Studi Roma Tre, via della Vasca Navale 84, 00146 Roma, Italy %10
\and Institute of Cosmology and Gravitation, Dennis Sciama Building, University of Portsmouth, Burnaby Road, Portsmouth, PO1 3FX %11
\and Astronomical Observatory of the University of Geneva, ch. d'Ecogia  16, 1290 Versoix, Switzerland%12
\and INAF - Osservatorio Astronomico di Trieste, via G. B. Tiepolo 11, 34143 Trieste, Italy %13
\and Institute for Astronomy, University of Edinburgh, Royal Observatory, Blackford Hill, Edinburgh EH9 3HJ, UK %14
\and Institute of Physics, Jan Kochanowski University, ul. Swietokrzyska 15, 25-406 Kielce, Poland %15
\and Department of Particle and Astrophysical Science, Nagoya University, Furo-cho, Chikusa-ku, 464-8602 Nagoya, Japan %16
\and Dipartimento di Fisica e Astronomia - Alma Mater Studiorum Universit\`{a} di Bologna, viale Berti Pichat 6/2, I-40127 Bologna, Italy %17
\and INFN, Sezione di Bologna, viale Berti Pichat 6/2, I-40127 Bologna, Italy %18
\and Institute d'Astrophysique de Paris, UMR7095 CNRS, Universit\'{e} Pierre et Marie Curie, 98 bis Boulevard Arago, 75014 Paris, France %19
\and Max-Planck-Institut f\"{u}r Extraterrestrische Physik, D-84571 Garching b. M\"{u}nchen, Germany %20
\and Laboratoire Lagrange, UMR7293, Universit\'e de Nice Sophia Antipolis, CNRS, Observatoire de la C\^ote d’Azur, 06300 Nice, France %21
\and Astronomical Observatory of the Jagiellonian University, Orla 171, 30-001 Cracow, Poland %22
\and National Centre for Nuclear Research, ul. Hoza 69, 00-681 Warszawa, Poland %23
\and Universit\"{a}tssternwarte M\"{u}nchen, Ludwig-Maximillians Universit\"{a}t, Scheinerstr. 1, D-81679 M\"{u}nchen, Germany %24
\and INAF - Istituto di Astrofisica Spaziale e Fisica Cosmica Bologna, via Gobetti 101, I-40129 Bologna, Italy %25
\and INAF - Istituto di Radioastronomia, via Gobetti 101, I-40129, Bologna, Italy %26
\and Dipartimento di Fisica, Universit\`a di Milano-Bicocca, P.zza della Scienza 3, I-20126 Milano, Italy %27
\and INFN, Sezione di Roma Tre, via della Vasca Navale 84, I-00146 Roma, Italy %28
\and INAF - Osservatorio Astronomico di Roma, via Frascati 33, I-00040 Monte Porzio Catone (RM), Italy %29
\and Institut Universitaire de France %30
\and Universit\'e de Toulon, CNRS, CPT, UMR 7332, 83957 La Garde, France %31
}

\offprints{First Author \\ \email{adam.hawken@brera.inaf.it}}
\authorrunning{A. ~J. ~Hawken et al.}
\abstract{We identified voids in the completed VIMOS Public Extragalactic Redshift Survey (VIPERS), using an algorithm based on searching for empty spheres. We measured the cross-correlation between the centres of voids and the complete galaxy catalogue. The cross-correlation function exhibits a clear anisotropy in both VIPERS fields (W1 and W4), which is characteristic of linear redshift space distortions. By measuring the projected cross-correlation and then deprojecting it we are able to estimate the undistorted cross-correlation function. We propose that given a sufficiently well measured cross-correlation function one should be able to measure the linear growth rate of structure by applying a simple linear Gaussian streaming model for the redshift space distortions (RSD). Our study of voids in 306 mock galaxy catalogues mimicking the VIPERS fields would suggest that VIPERS is capable of measuring $\beta$ with an error of around $25\%$. Applying our method to the VIPERS data, we find a value for the redshift space distortion parameter, $\beta = 0.423^{+0.104}_{-0.108}$, which given the bias of the galaxy population we use gives a linear growth rate of $f\sigma_8 = 0.296^{+0.075}_{-0.078}$ at $z = 0.727$. These results are consistent with values observed in parallel VIPERS analysis using standard techniques.}
\maketitle
\section{Introduction}
Different cosmological models, and different theories of gravity, predict that the large scale distribution of matter should be structured in subtly different ways. The light emitted from galaxies can be used as a proxy to trace this weblike structure. The cosmic web can be split into different component structures that show different properties, namely nodes (clusters), filaments, walls, and voids. Cosmic voids are the most underdense regions of the universe, and compose most of its volume \citep{2004MNRAS.350..517S}. Their abundance can be used as a probe of the growth of structure \citep{2013MNRAS.434.2167J}. They are also the most dark energy dominated environments and so are ideal places in which to study the vacuum energy and to search for signatures of modified gravity \citep{2004ApJ...605....1G, 2013MNRAS.431..749C, 2015MNRAS.451.4215Z}. 

There are many competing explanations for the observed accelerating expansion of the Universe. Many of these models can reproduce the same expansion history, so measurements of the expansion history alone (either using standard candles like type 1a supernovae, or standard rulers like Baryon Acoustic Oscillations) cannot discriminate between them. However, theories that modify general relativity or the equation of state of dark energy may alter the effective strength of gravity and thus also the growth rate of structure. Therefore measuring the growth rate of structure at different redshifts is necessary to break the degeneracy between modified gravity and dark energy \citep{2009arXiv0901.0721A}.

Galaxies that trace cosmic structure are subject to motions in addition to the Hubble flow. These motions contribute to the observed redshift of a galaxy and distort its apparent position in space \citep{1987MNRAS.227....1K, 1998ASSL..231..185H}. 
Actually measuring the growth rate of structure is a technical challenge because even on the largest scales accessible to cosmological surveys the gravitational peculiar motions of galaxies are not fully linear.
But the density of material close to the edges of voids is the same order of magnitude as the mean cosmic density. Therefore the relationship between density and velocity fields should be linear. 
Here we propose a novel method that utilises the linear nature of the velocity field around cosmic voids to extract a measurement of the growth rate of structure.

A galaxy in or close to the edge of a void is probably being evacuated away from the void centre, falling onto the surrounding structure under the influence of gravity \citep{padillaetal05, 1993ApJ...410..458D}. These redshift space distortions (RSD) introduce an anisotropy to the void-galaxy cross-correlation function, $\xi_{vg}$ \citep{pazetal13, 2014JCAP...12..013H, 2015arXiv150704363H, 2016arXiv160201784H, 2016arXiv160305184C, 2016arXiv160505352C, 2016arXiv160603092A}. If all anisotropy in the void-galaxy cross-correlation function arises via RSD, and the relationship between the velocity and density fields is understood, then the strength of the RSD signal can be measured given a model for the isotropic density field around voids.

In Section \ref{sec:voidsinvipers} we give an overview of the search for voids in our data set, the VIMOS Public Extragalactic Redshift Survey (VIPERS\footnote{\tt http://vipers.inaf.it/}). We also describe the mock catalogues used in our analysis. Section \ref{sec:model} describes a toy model for the void-galaxy cross-correlation function, which we shall later use to test our methodology. Section \ref{sec:rsd} outlines our model for the anisotropies caused by linear redshift space distortions. Our measurements of the cross-correlation are described in Section \ref{sec:measurements}. Section \ref{sec:deproj} describes how by deprojecting the projected void-galaxy cross-correlation function we can estimate the realspace void density profiles. Section \ref{sec:growthrate} describes how we built covariance matrices from the mock catalogues and fit our model to the mocks and subsequently to the data in Section \ref{sec:vipersgrowth}. By doing this it is possible to extract a measurement of the growth rate of structure, $f(\Omega)$. We conclude in Section \ref{sec:conclusion}, where we also discuss our results and methodology, with reference to recent progress by others in this field.

\section{The search for voids in VIPERS}
\label{sec:voidsinvipers}

The VIMOS Public Extragalactic Redshift Survey (VIPERS\footnote{\tt http://vipers.inaf.it/}) is an ESO Large Programme, started at the end of 2008, to map in detail the spatial distribution of galaxies, with magnitude $i_{AB}<22.5$, over an unprecedented volume of the Universe up to $z\sim1$.  Its goals are to accurately and robustly measure galaxy clustering, galaxy properties, and the growth of structure at an epoch when the Universe was about half its current age. The galaxy target sample is based on 5-band photometric data from the Canada-France-Hawaii Telescope Legacy Survey Wide catalogue \citep[CFHTLS-Wide;][]{2012SPIE.8448E..0MC}. VIPERS is split over two CFHTLS fields named W1 and W4.
%The VIPERS large scale structure catalogue, which is now complete, contains 50829 objects in W1 and 26377 objects in W4. 

The survey is particularly narrow in declination ($1.8^\circ$ in W1 and $1.6^\circ$ in W4) which makes it difficult to use common void finding techniques such as the watershed algorithm \citep{Platenetal2007, 2008MNRAS.386.2101N, sutteretal12}. We therefore developed an algorithm that searches for voids using empty spheres, which is described in detail in \citet{2014A&A...570A.106M}.

Following \citet{2014A&A...570A.106M} we searched for voids in a volume limited sample of galaxies from the VIPERS final data release with a redshift $0.55 < z < 0.9$, selecting galaxies with an absolute magnitude $M_B - 5{\rm log}h < -19.3 - z$, that have spectroscopic flags $\geq2$. 
This corresponds to regions approximately $695\mpcoh$ long, and 58 by $265 \mpcoh$ in W1, and 51 by $168 \mpcoh$ in W4 (at a redshift of $z=0.75$). The total volume in which we search for voids is then approximately $1.6\times 10^{7} (h^{-1}{\rm Mpc})^3$. Our volume limited catalogue for W1 contains 23210 objects and for W4 contains 11426 objects. %
We then grow empty spheres on a fine regular grid of resolution $0.7\mpcoh$. 
The VIMOS mask leaves gaps corresponding to $\sim1-2\mpcoh$, to avoid selecting spurious under densities generated by masking effects we limit ourselves to only searching for the most significant empty spheres. In practice this means that the empty spheres we are interested in have a radius $\gtrsim 8 \mpcoh$. This is smaller than the minimum radius in  \citet{2014A&A...570A.106M}, which was defined in a different way and was overly conservative. Spheres are discarded if more than $20\%$ of their volume lies outside the survey boundaries. We define voids as being statistically significant spheres that do not overlap.
 We identified 822 voids in the W1 field of VIPERS, and 441 voids in W4. 

\begin{figure}
\begin{center}
\includegraphics[width=0.5\textwidth]{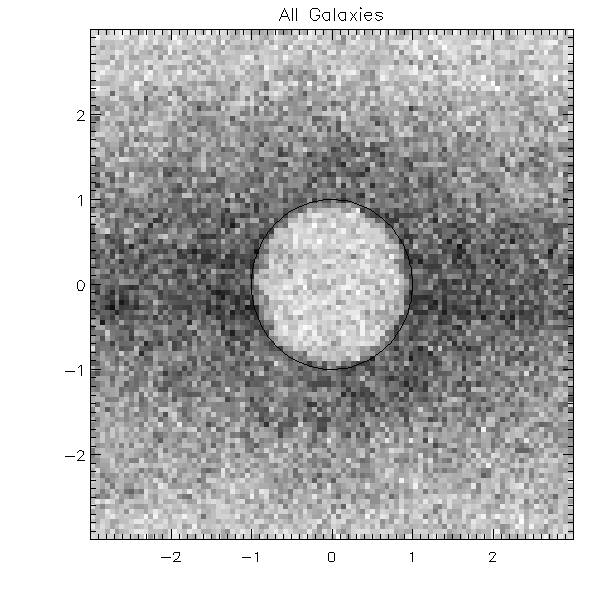}
\caption{Stacked voids in VIPERS PDR-2. This figure shows the density of galaxies in PDR-2 relative to the centres of voids. The $x-y$ plane of the figure corresponds to the plane of the sky in comoving coordinates, rescaled to the radii of the voids.  The black circle indicates $r/r_s = 1$, i.e. the normalised radius of the stacked voids. The thickness of each slice in the stack is 0.25 void radii.}
\label{fig:voidstack}
\end{center}
\end{figure}

Figure \ref{fig:voidstack} shows all the voids in the two fields stacked on top of one another. The $x-y$ plane of Figure \ref{fig:voidstack} corresponds to the plane of the sky in comoving coordinates, rescaled to the radii of the voids. The thickness of each slice in the stack is 0.25 void radii. The points represent the density of galaxy positions, which have been rescaled by the radii of the spheres. One can see that on average these under densities are spherically symmetric with an apparent overdense ridge between one and two void radii from the centre. One can also see that there is an enhancement in the apparent density of galaxies along the $x$ axis: this is a systematic effect due to the geometry of VIPERS. The two fields are broad in right ascension and narrow in declination. This has the effect that galaxies are more likely to be found to the left or right of voids on the plane of the sky than above or below. Systematic effects such as this caused by the geometry of the survey are the primary reason why the stacked density profile is not as useful a measurement as the void-galaxy cross-correlation function.

\begin{figure}
\begin{center}
\includegraphics[width=0.5\textwidth]{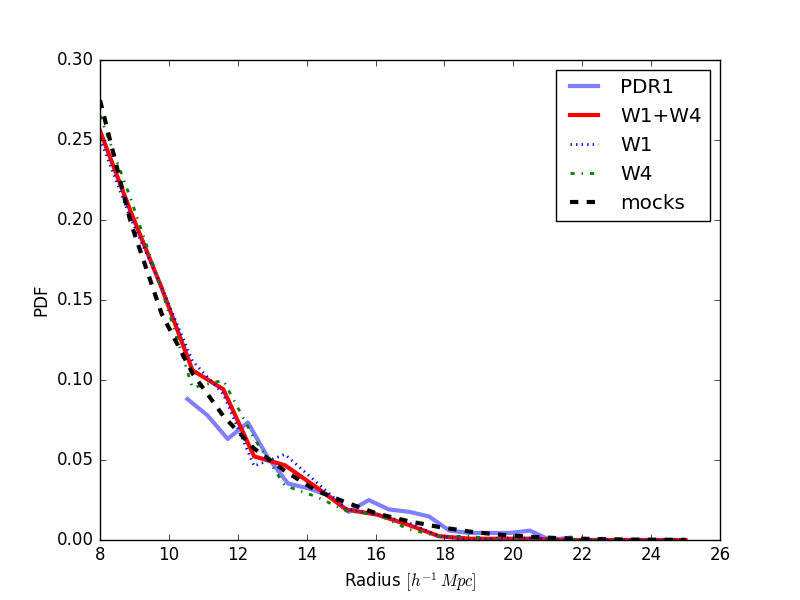} 
\caption{The normalised histogram of void radii in this data set (red solid line) compared to those in the mock catalogues (black dashed line). The blue and green dotted lines show the two individual fields in this data set. The distribution of void sizes in the data is in very good agreement with the mocks. The histogram of void sizes in \citet{2014A&A...570A.106M} is also plotted (blue solid line). Note that this has been renormalised to account for the change in minimum void radius is this work.}
\label{fig:size_hist}
\end{center}
\end{figure}

There is a notable increase in sky coverage, mainly in W1, in PDR-2 compared to the first public data release (PDR-1). Additionally, pointings within the survey borders that were missing in PDR-1 have since been reobserved. This has had an effect on the apparent size and distribution of voids near these regions. 
Although there is not a one to one correspondence between voids in the current data set and those in PDR-1, in general the properties of the voids in the new catalogue are not appreciably different from those presented in \citet{2014A&A...570A.106M}. Figure \ref{fig:size_hist} shows the normalised histogram of void radii in this data set (red solid line) compared to the mock catalogues (black dashed line) (see Section \ref{sec:mocks} for a description of the mocks). The distribution of sizes is consistent with the mock catalogues and PDR-1 (solid blue line). There are no suspicious differences between the two fields (blue and green dotted lines).

\subsection{Mock galaxy catalogues}
\label{sec:mocks}
The mock galaxy catalogues we have used have been constructed by populating a large N-body simulation with galaxies using a Halo Occupation Distribution (HOD). The haloes were taken from the dark matter halo catalogue of the BigMultiDark simulation \citep{Pradaeetal12}. This simulation has a $\Lambda$CDM cosmology ($\Omega_m = 0.31$, $\Omega_\Lambda = 0.69$, $\Omega_b = 0.048$, $\sigma_8 = 0.82$, $n_s = 0.96$, $h = 0.7$). The original halo catalogue is limited in mass to haloes below $\sim10^{12} M_\odot \,h^{-1}$ due to the mass resolution of the simulation. In order to produce mock galaxies as faint as those in VIPERS the simulation was first repopulated with haloes of masses below the resolution limit by reconstructing the density field from the dark matter field, following the method described in \citet{2013MNRAS.435..743D}. The haloes were then populated using the HOD, for which the redshift evolution was calibrated using clustering measurements from VIPERS. A full description of method and parameters can be found in \citet{delatorreetal13}, and in the parallel paper \citet{DeLaTorre2016}. 

Mocks were then extracted from the catalogue, using a VIPERS-like colour selection and magnitude limit, $i_{\rm AB} < 22.5$. The selection function, $n(z)$, in these parent mocks was then explicitly matched to the observed redshift distribution of galaxies in the two VIPERS fields combined. Gaussian errors on redshifts were then applied, $\sigma_v = 135 \kms$, corresponding to the velue estimated in PDR-1. Spectroscopic masks were built for each mock using the slit positioning software, SSPOC \citep{2004astro.ph..9252B}. The target sampling rate (TSR), introduced by SSPOC, is a function of the local surface density of galaxies. Thus the TSR values of the mocks differ slightly from those in the real data. 

Furthermore, not all measurements of spectra using VIMOS are successful, so the Spectroscopic Success Rate (SSR) varies from quadrant to quadrant. The SSR depends on a number of factors such as the seeing on the night the observations were taken, distance of the pointing from the ecliptic plane, and the magnitude of the source. To account for this we have randomly down-sampled the mocks to have the same density as the VIPERS data.

\section{Modelling the void-galaxy cross-correlation function}
\label{sec:model}

In this section we describe a simple model for the void-galaxy cross-correlation function, $\xi_{vg}$.

The integrated density contrast in a void-centred sphere of radius, $r$, and volume, $V$, is
\begin{equation}
\Delta(r) = \frac{1}{V} \int_V \bigg( \frac{\rho(r)}{\bar{\rho}} - 1\bigg) \; {\rm d} V.
\end{equation}
The void galaxy cross-correlation function is defined as 
\begin{eqnarray}
\xi_{vg}(r) &=& \frac{\rho(r)}{\bar{\rho}} - 1, \\
 &=& \delta_g(r),
\end{eqnarray}
where $r$ is the distance from the void centre \citep{1980lssu.book.....P}. Thus the void galaxy cross-correlation function can be expressed in terms of the integrated void density profile,
\begin{equation}
\label{eqn:xi_Delta}
\xi_{vg}(r) = \frac{1}{3 r^2}\frac{\rm d}{{\rm d}r} \big(r^3 \Delta(r)\big).
\end{equation}

There are several proposed functional forms for the void density profile in the literature. These can broadly be divided into two categories: phenomenological models that seek to fit the functional form of the void density profile \citep[e.g.][]{2014PhRvL.112y1302H, pazetal13, 2015MNRAS.449.3997N}, and theoretically motivated models \citep[e.g.][]{2014arXiv1405.1555F}. Some of these models include a free parameter that allows for an overcompensating ridge around the void. Objects with ridges like this tend to be smaller voids embedded inside overdensities and are actually contracting, being crushed by the surrounding overdensity \citep{2004MNRAS.350..517S}. Velocities in the vicinity of such objects may be far from linear.

\citet{2004MNRAS.350..517S} first observed that voids can be divided into two populations based on environment. Void-in-void objects are embedded in underdense regions. These voids tend to be larger, and behave in a very linear way, expanding as structure in the Universe grows. The density profiles of these voids typically asymptote to the mean density of the Universe with little or no compensating ridge around them. Void-in-cloud objects are voids that are embedded in overdense regions. These voids typically have heavily or overcompensated density profiles and their dynamical properties are less linear. Furthermore, they typically shrink as structure grows, becoming crushed by the surrounding overdensity. However, the interiors are still being evacuated and their immediate surroundings are still expected to be linear.

\begin{figure}
\begin{center}
\includegraphics[width=0.5\textwidth]{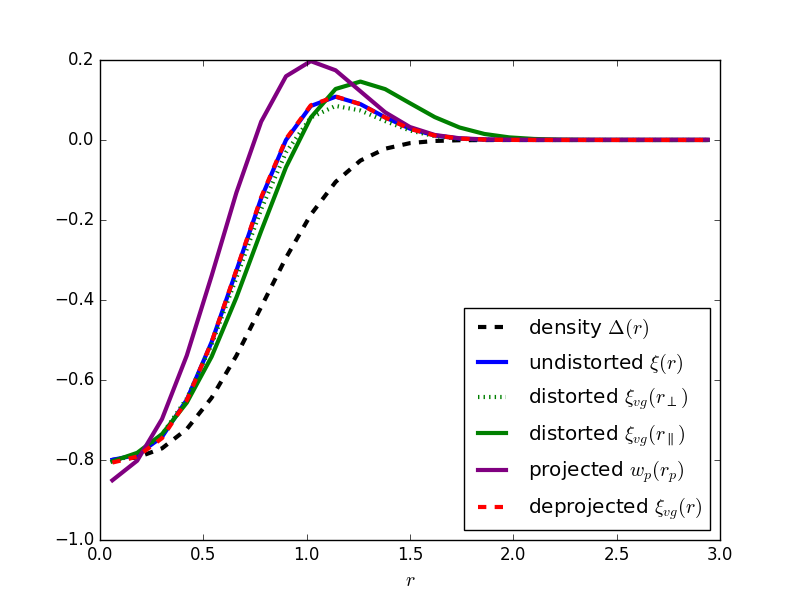} 
\caption{Model for the void-galaxy cross-correlation function. The integrated density contrast, Equation (\ref{eqn:Delta}), is plotted as a black dashed line. The one dimensional void-galaxy cross-correlation function, without redshift space distortions, is plotted as a solid blue line. The void-galaxy cross-correlation function with redshift space distortions, as seen directly along the line of sight, is plotted as a solid green line. The dotted green line is the same model as seen tangential to the line of sight. The projected cross-correlation function is plotted as a purple line. The deprojection is plotted as a red dashed line, it matches the blue line very closely. The values of the model parameters are $\beta=0.8, \sigma_v=13.4 \, {\rm km \, s}^{-1} \,(\mpcoh)^{-1}, \delta_c=-0.8,r_v=0.9,\alpha=3.0$.}
\label{fig:1dmodel}
\end{center}
\end{figure}

Here we propose a simple stretched exponential form for the integrated density contrast of galaxies,
\begin{equation}\label{eqn:Delta}
\Delta(r) = \delta_c \exp \bigg(-\bigg(\frac{r}{r_v}\bigg)^\alpha\bigg).
\end{equation}
This model has three parameters: the central density of the void, $\delta_c$; some scale radius, $r_v$; and the shape parameter, $\alpha$. The correlation function for this profile is easy to write analytically:
\begin{equation}
\xi_{vg}(r) = \delta_c\bigg(1 - \frac{\alpha}{3}\bigg(\frac{r}{r_v}\bigg)^\alpha\bigg)\exp\bigg(-\bigg(\frac{r}{r_v}\bigg)^\alpha\bigg).
\end{equation}
This simple functional form is plotted in Figure \ref{fig:1dmodel}. It is interesting to note that a Gaussian profile is a special case of this model, where $\alpha = 2$.  We shall use this model density profile to test our method for measuring the growth rate in Section \ref{sec:tst_tm}, but we shall not be fitting it to the observed density profile in this paper.

\section{Linear redshift space distortion model}
\label{sec:rsd}
In this section we describe our linear model for the redshift space distortions around voids.
The line of sight pairwise velocity distribution can generally be described using the streaming model, so the anisotropic void-galaxy cross-correlation function can be written:
\begin{equation}
\label{eqn:xi_aniso}
\begin{split}
1 + \xi_{vg}(r_\parallel, r_\perp) = \int_{-\infty}^{+\infty} & \frac{{\rm d}w_3}{\sqrt{2\pi}\sigma_v(r)} \times \\
& \exp \bigg(-\frac{(w_3 - v(r){r_3 / r})^2}{2\sigma_v^2(r)} \bigg) [ 1+\xi_{vg}(r)] ,
\end{split}
\end{equation}
where $r_3 = r_\parallel -w_3/H_0$, $r^2 = r_\perp^2 + r_3^2$, and $w_3$ is the line of sight component of the pairwise velocity.

The velocity dispersion of galaxies, $\sigma_v(r)$, is a function of distance from the void centre and has units of $\kms \mpcoh$, i.e. velocity per void radius. Attempts have been made to study $\sigma_v(r)$ in simulations and concluded that its functional dependence on the separation of void-galaxy pairs, and on the local matter density, is not well constrained \citep{2015arXiv150704363H}. 

A known and quantifiable source of apparent dispersion in the streaming velocity of galaxies is the error on the redshift measurement. In our mock galaxy catalogues a Gaussian error of $\sigma_z=135 \, {\rm km \, s}^{-1}$ was applied. This is actually a bit small compared to the estimated error in this data set, $\sigma_z=140\, {\rm km \, s}^{-1}$. By weighting using the distribution of void sizes we can calculate the effective contribution to $\sigma_z$ ,
\begin{equation}
\sigma_v = \sum_i \frac{\sigma_z}{r_s^i}w_i,
\end{equation}
where $r_s^i$ is the radius of voids in bin $i$ and $w_i$ is the weight of that bin. The weights are determined using the histogram of void sizes (see Figure \ref{fig:size_hist}), normalised such that $\sum_i w_i = 1$. For the mocks the effective dispersion is $\sigma_v=13.4\, {\rm km \, s}^{-1} \,(\mpcoh)^{-1}$, and for the data this is $\sigma_v = 13.8\, {\rm km \, s}^{-1} \,(\mpcoh)^{-1}$.

Because the densities involved are very low, the gravitational dynamics of galaxies around voids, particularly larger ones, remain in the linear regime \citep{2016arXiv160305184C}.
This should be particularly true for our void sample because our voids are relatively large and so are expected to be more linear.
Close to the centres of voids $\delta \sim -1$, so the relationship between the density and velocity fields is not strictly speaking linear. However, because these regions are very sparsely populated by tracers they do not contribute much to the overall signal and so their non-linear contributions can be ignored.
We therefore make the assumption that the linear estimate for the relationship between the density and velocity fields remains valid, and that the relationship between the velocities of galaxies and that of matter is unbiased \citep{1980lssu.book.....P},
\begin{equation}
\label{eqn:vel}
v(r) = - \frac{H(z)}{1+z }r \Delta(r) \frac{\beta}{3},
\end{equation}
where  $\beta=f(z)/b$ is the redshift space distortion parameter, with $b$ being the galaxy bias and $f(z)$ the linear growth rate parameter, defined as the logarithmic derivative of the linear growth factor, $D(a)$, with respect to the scale factor, $f=d \ln D/d \ln a$.
The growth factor is commonly parameterised as $f(z) = \Omega_m^\gamma(z)$, which is useful because it gives a useful approximate solution to the growth equation for a wide variety of gravity models \citep{1980lssu.book.....P, 1998ApJ...508..483W, 2005PhRvD..72d3529L, 2007APh....28..481L}. In standard general relativity $\gamma \approx 0.55$; any deviation from this value could be taken as evidence in favour of modifying general relativity. 

A correct description of the velocity field should also consider the impact of galaxy biasing. It is well known that galaxies inhabiting voids have notably different properties from galaxies outside of voids, and in fact this is the subject of many void studies. These studies have established that galaxies inhabiting voids are typically bluer, of later type, and with higher specific star formation rate than other field galaxies \citep{2004ApJ...617...50R, 2006MNRAS.372.1710P, 2007arXiv0710.2783V, 2012MNRAS.426.3041H, 2012AJ....144...16K, 2014MNRAS.445.4045R}. Thus one should expect that the galaxy bias in this case is heavily scale dependent. Models have been proposed to describe how haloes are biased as a function of distance from the void centre \citep{2014MNRAS.441..646N}, so extending the model to include a scale dependent bias would certainly be possible. However, for now, we consider the bias to be constant.

We also make the assumption that the Hubble expansion rate and the angular diameter distance are well constrained and therefore we neglect any potential geometric distortions due to the Alcock-Paczynski effect.

\section{Measuring the void-galaxy cross-correlation function}
\label{sec:measurements}

In this section we describe our estimator for the void-galaxy cross-correlation function, $\xi_{vg}$. The estimated value of $\xi_{vg}$, in some bin of separation $ij$, is equal to the estimated overdensity in that bin,
\begin{eqnarray}
\hat{\xi}_{vg}(r_\parallel^i,r_\perp^j) &=& \hat{\delta}_g^{ij} \\
&=& \frac{n_g^{ij}}{f^{ij} \bar{n}_g} - 1,
\end{eqnarray} 
where $\bar{n}_g$ is the mean number density of galaxies per bin, $n^{ij}_g$ is the number of galaxies counted in bin $ij$, and $f^{ij}$ is the fraction of the bin which is unmasked, i.e. which lies completely within the survey boundaries. $f^{ij}$ is estimated using a random catalogue with the same angular and redshift selection function as the galaxies, $f^{ij} = n^{ij}_r / \bar{n}_r$, where $n^{ij}_r$ is the number of random points counted in the bin and $\bar{n}_r$ is the mean number density of random points. The estimator of the cross-correlation can then be written
\begin{equation}
\hat{\xi}_{vg}(r_\parallel^i,r_\perp^j) = \frac{n^{ij}_g}{n^{ij}_r} \frac{N_r}{N_g} -1,
\end{equation}
where $N_r$ is the total number of random points and $N_g$ is the total number of galaxies. This is just the Davis and Peebles estimator for the cross-correlation \citep{davispeebles83}.

As mentioned above, random catalogues were constructed in such a way as to have the same angular and radial selection functions as the data. We did this by applying the same photometric masks to initially uniform distributions of random points covering the two fields. Redshifts were then assigned to the random points by sampling from the redshift distribution of mock galaxies.

The cross-correlation function presented here is the cross-correlation between the centres of the maximal spheres and the full VIPERS PDR-2 galaxy catalogue. Void-galaxy pair separations are scaled in units of the radius of the maximal spheres, $r_s$, so that $\xi_{vg}(\tilde{r}_\parallel, \tilde{r}_\perp) = \xi_{vg}(r_\parallel/r_s, r_\perp/r_s)$. 

\begin{figure*}
\begin{center}
\includegraphics[width=\textwidth]{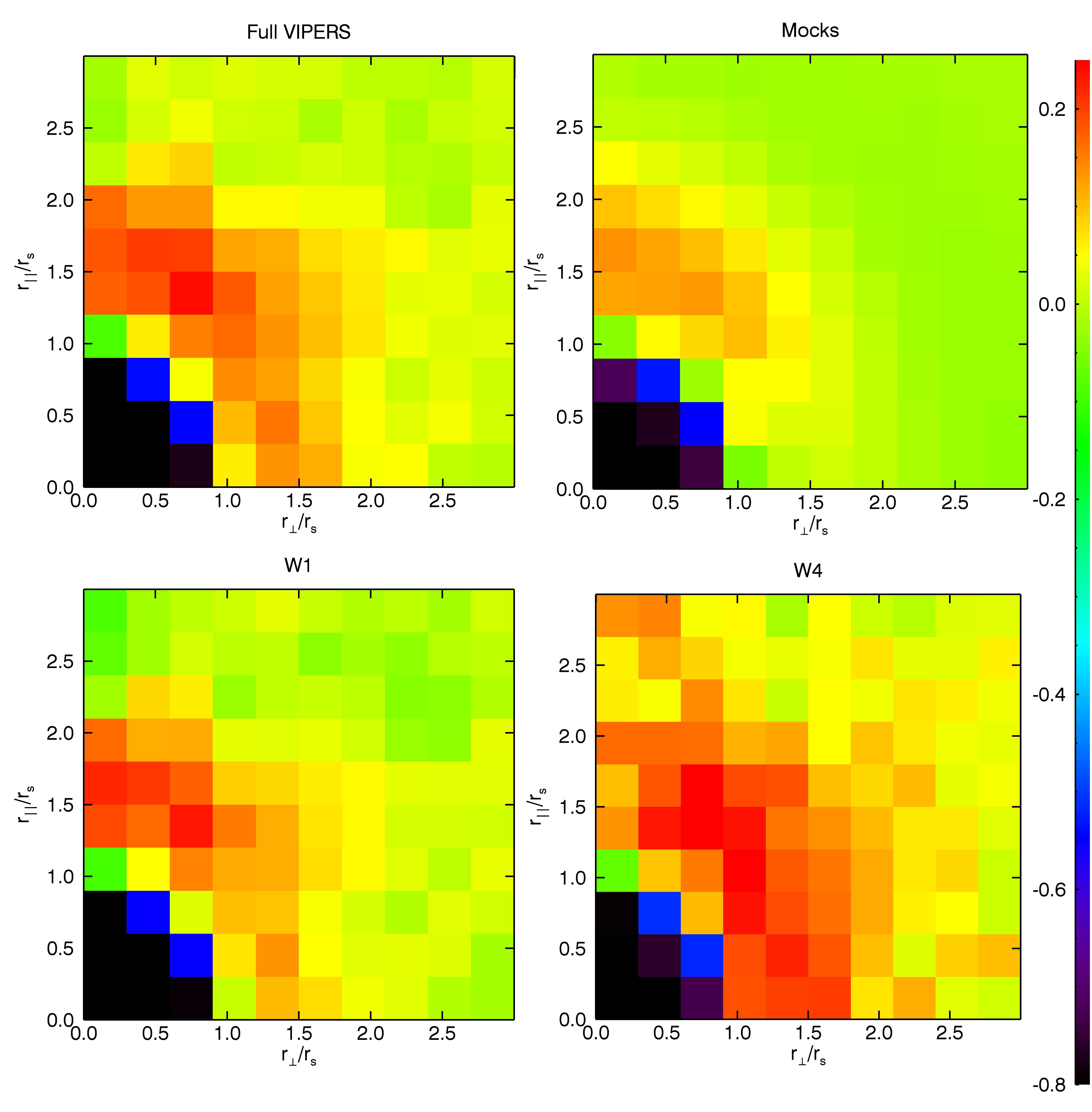} 
\caption{cross-correlation function between the centres of voids and the full sample of galaxies in VIPERS. The bottom two panels show the measured cross-correlation in the two individual VIPERS fields. The top left panel shows the average of these two fields. The top right panel shows the mean cross-correlation function of the 306 mock catalogues for comparison. The axes are in units of void radii.}
\label{fig:cc_all}
\end{center}
\end{figure*}

Figure \ref{fig:cc_all} shows $\xi_{vg}$ measured in $10\times10$ bins individually in the two separate VIPERS fields, and the combined measurement of the full sample. The enhancement of the correlation function along the line of sight is clearly visible. The measurement in the W4 field appears to be noisier than W1, but this is to be expected because the field is smaller. For comparison we also plot the mean cross-correlation of the 306 mock catalogues.

\section{Deprojecting the cross-correlation}
\label{sec:deproj}

In order to determine the degree to which the anisotropic cross-correlation function is distorted, we must first seek to determine what the undistorted cross-correlation looks like. We do this by deprojecting the projected cross-correlation function \citep{2003ApJ...586..718E, 2007MNRAS.381..573R, 2014MNRAS.443.3238P}.

By integrating along the line of sight direction we can obtain a measurement of the projected void-galaxy cross-correlation function,
\begin{equation}
w_{vg}(r_p)=2\int_0^\infty \xi_{vg}(r_\perp,r_\parallel){\rm d}r_\parallel.
\end{equation}
The projected cross-correlation is, in principle, unaffected by redshift space distortions. In practice, this integral does not extend to infinity but to some $r_\parallel^{\rm max}$, which is constrained by the depth of the survey. Because $\xi_{vg}(r)$ is expected to be zero at large $r$, we truncate the integral at $r_\parallel^{\rm max}/r_v = 3$. Truncating at larger distances than this simply adds noise to the measurement. 
The projected void galaxy cross-correlation function can also be written as
\begin{equation}
\label{eqn:wvg}
w_{vg}(r_p) = 2 \int_{r_p}^\infty \xi_{vg}(r) \frac{r {\rm d}r}{\sqrt{r^2 - r_p^2}} \; .
\end{equation}
Given that we assume the true cross-correlation function to be isotropic we can invert Equation (\ref{eqn:wvg}) using the Abel transform to obtain an estimate of $\xi_{vg}(r)$,
\begin{equation}
\xi_{vg}(r) = -\frac{1}{\pi}\int_r^\infty \frac{{\rm d}w(r_p)}{{\rm d}r_p}\frac{{\rm d}r_p}{\sqrt{r_p^2 - r^2}}.
\end{equation}
For a given bin $r_i$ this can be calculated using
\begin{equation}
\label{eqn:deproj_numerical}
\xi_{vg} (r_i) = -\frac{1}{\pi}\sum_{j\geq i} \frac{w_{vg, j+1} - w_{vg, j}}{r_{p,j+1} - r_{p,j}}\ln \left(\frac{r_{p, j+1} + \sqrt{r_{p,j+1}^2 -  r_{i}^2}}{r_{p,j} + \sqrt{r_{p,j}^2 -  r_{i}^2} }\right),
\end{equation}
where $w_{vg, j}$ is the value of $w_{vg}(r_{p,j})$, the projected cross-correlation function in bin $r_{p,j}$. 
The number of bins will have an effect on the accuracy of the projection and deprojection of the correlation function. Firstly by introducing integration noise when integrating over the line of sight. Secondly because when applying the model of the RSD we linearly interpolate both $\xi(r)$ and $\Delta(r)$. Thirdly because deprojecting involves numerical differentiation. In practice we can reduce any systematic bias introduced by the numerical differentiation in Equation (\ref{eqn:deproj_numerical}) by interpolating between bin centres using a cubic spline. 

The number of bins in which we can measure \smash{$\xi_{vg}$} is limited not only by the amount and quality of the data but also by the number of mocks we have available to build the covariance matrices. When we measure $\xi_{vg}(r_\parallel,r_\perp)$ in $25\times 25$ bins in the data it is very noisy. However, integrating over $r_\parallel$ removes much this noise. Therefore we measure the projected correlation function in $25\times25$ bins, but when we deproject and then use the result in an anisotropic fit to $\xi_{vg}(r_\parallel,r_\perp)$ we fit to $10 \times 10$ bins (as shown in Figure \ref{fig:cc_all}).

To obtain the empirical estimate of the void density profile we first combine the measured cross-correlation functions in the two fields by weighting them based on the number of voids found in that field,
\begin{equation}
\xi_{W1+W4} = \xi_{W1}\frac{N_{\rm voids}^{W1}}{N_{\rm voids}^{\rm tot}} + \xi_{W4}\frac{N_{\rm voids}^{W4}}{N_{\rm voids}^{\rm tot}}.
\end{equation}
The deprojection procedure can then be followed to build an estimate of the undistorted $\xi_{vg}$ based on all the available data. Because there is no reason to believe that the density profiles of voids in the two fields would be significantly different, we can apply the same model to both fields. This also allows us to make a meaningful comparison between measurements of the growth rate from the two fields.

\section{Measuring the growth rate}
\label{sec:growthrate}

This section describes our method for constraining the growth rate of structure by fitting the model outlined in Section \ref{sec:rsd} to the measurement of the void-galaxy cross-correlation function, $\xi_{vg}(r_\parallel, r_\perp)$, presented in Section \ref{sec:measurements}. 

We measured $\xi_{vg}$ in 306 mock galaxy catalogues covering W1 and W4. From these measurements we constructed covariance matrices for each field, Section \ref{sec:covmat}.
The input cosmology of the mocks is known, and thus so is the linear growth rate $f(z)$. However, our method provides us with an estimate of $\beta = f/b$, and so to confirm that we are able to constrain the growth rate correctly we must first measure the bias of the galaxies we are using in the mocks, Section \ref{sec:bias}. Once the correct growth rate has been extracted from the mocks, \ref{sec:mock_tests} , and any systematic bias in the measurement quantified, we can place a constraint on the growth rate in the data using the variance of recovered values from the mocks as our error bar, Section \ref{sec:vipersgrowth}.

\subsection{Covariance matrix and likelihood estimation}
\label{sec:covmat}

We ran our void finding algorithm on each of the mocks and measured the void-galaxy cross-correlation function $\xi_{vg}$ in order to construct a covariance matrix. 
%Redshift space distortions move galaxies from one $r_\parallel, r_\perp$ bin to another, leading to
There is a strong covariance between bins, this makes the covariance matrix highly non-diagonal. Thus it is important that the full covariance matrix is used to constrain the parameters of the model and not just the variance of the individual bins. 

 An important point to note is that in this experiment \smash{$\xi^{\rm model}_{vg}$ is built} using the observed cross-correlation, \smash{$\xi^{\rm obs}_{vg}$}, and is therefore not independent of the data. Noise present in the observations propagates through to noise in the model. Failing to account for this propagation of noise leads to a biased estimate of the growth rate and an overestimation of the error. However, if we take care to use the correct covariance matrix and to apply the appropriate Bayesian correction factors to it then we can mitigate any introduced biases to recover the correct parameter values and their uncertainty.

$\pmb{\Delta}$ is a matrix defined as the difference between the observed anisotropic void galaxy cross correlation function and the reprojected cross correlation given a model for the RSD,
\begin{equation}
\Delta_i = \xi_{vg}^{\rm obs}(r_\parallel,r_\perp)_{i} - \xi_{vg}^{\rm model}(r_\parallel,r_\perp)_{i},
\end{equation}
where $i$ indicates the bin in $r_\parallel$ and  $r_\perp$. The mean residual between the model, given the fiducial cosmology, and $\xi_{vg}(r_\parallel,r_\perp)$ observed in the mocks is 
\begin{equation}
\label{eqn:mu}
\mu_i = \frac{1}{N_{\rm mocks}}\sum_{k=1}^{N_{\rm mocks}} \Delta_i^k.
\end{equation}
This quantifies the extent to which the model is biased.
The expectation value of the data does not correspond to the model and so $\mu \neq 0$. 
This is because our model for the RSD is an imperfect description of the anisotropy, so even if the cosmology is known then the exact anisotropic cross correlation cannot be completely recovered. One consequence of this is that the expectation value of the data matrix is not equal to the true covariance matrix. i.e. that $\langle \pmb{\Delta \Delta} \rangle \neq \pmb{C}$. The correct covariance matrix in this instance can be defined as the expectation of the difference between the model and the observations minus the mean residual,
\begin{equation}
\label{eqn:covmat}
C_{ij} = \frac{1}{N_{\rm mock} - 1} \sum_{k=1}^{N_{\rm mock}} (\Delta_{i}^k - \mu_i) (\Delta_j^k - \mu_j ).
\end{equation}
The likelihood of a set of parameter values, $\pmb{\theta}$, given the observation is then,
\begin{equation}
\mathcal{L}(\pmb{\theta}) = \exp\bigg({\frac{-\chi^2}{2}}\bigg),
\end{equation}
where
\begin{equation}
\chi^2 = (\pmb{\Delta} - \pmb{\mu})^T \pmb{C} (\pmb{\Delta} - \pmb{\mu}),
\end{equation}
with $\pmb{\mu}$ being the residual matrix as measured in the mocks, given by Equation \ref{eqn:mu}. This assumes that the likelihood $\mathcal{L}(\pmb{\theta})$ is Gaussian, which we do not know to be true. We can test the Gaussianity of the likelihood by looking at the scatter of recovered values from mock catalogues (see Section \ref{sec:mock_tests}).
The covariance matrices are calculated individually for each field. The combined likelihood for the full survey is calculated by summing the $\chi^2$ for each field:
\begin{equation}
\chi^2_{W1+W4} =  (\pmb{\Delta}_{W1} - \pmb{\mu})^T \pmb{C}_{W1} (\pmb{\Delta}_{W1} - \pmb{\mu}) + (\pmb{\Delta}_{W4} - \pmb{\mu})^T \pmb{C}_{W4} (\pmb{\Delta}_{W4} - \pmb{\mu}).
\end{equation}

The covariance matrix defined in Equation (\ref{eqn:covmat}) is biased because the number of mocks used to produce it is finite, and of the same order as the number of degrees of freedom. The bias of this estimate can be corrected for by replacing it in the likelihood calculation with a matrix $\pmb{\Psi}$ defined as \citep{2007A&A...464..399H}
\begin{equation}
\pmb{\Psi} = (1-\mathcal{D}) \, \pmb{C}^{-1},
\end{equation}
where 
\begin{equation}
\mathcal{D} = \frac{N_{\rm bins} + 1}{N_{\rm mocks} - 1}.
\end{equation}
Note that we do not incorporate the remaining statistical uncertainty in $\pmb{C}$ into our likelihood, although in principle this can be done \citep{2016MNRAS.456L.132S}.

The mock catalogues were built using an HOD which was constructed so that the projected two point clustering of galaxies matched observations. They were not constructed with an analysis of void properties in mind. 
Furthermore, %not all of the mock galaxy catalogues used to produce the covariance matrix are independent. 
regions corresponding to W1 and W4 were sometimes cut from the same simulation boxes. Additionally the bias and colour evolution of galaxies in the mocks are not completely accurate. These effects can lead to inaccuracies of our covariance matrix. These errors in the covariance matrix should be propagated correctly.
 
We wish to determine the combined error on the measurement, including both the uncertainties inherent in the data and the noisy covariance matrix. In order to obtain an unbiased estimate of the full error we must also multiply the inverse covariance matrix by a factor of $m_1$ \citep{2014MNRAS.439.2531P},
\begin{equation}
m_1 = \frac{1+B(N_{\rm bins} - N_p)}{1+A+B(N_p+1)},
\end{equation}
where $N_p$ is the number of parameters in the model, and where
\begin{eqnarray}
A &=& \frac{2}{(N_{\rm mocks} - N_{\rm bins} -1)(N_{\rm mocks} - N_{\rm bins} - 4)}, \\
B &=& \frac{N_{\rm mocks}-N_{\rm bins}-2}{(N_{\rm mocks} - N_{\rm bins} - 1)(N_{\rm mocks} - N_{\rm bins} - 4)}.
\end{eqnarray}

An accurate estimate of the uncertainty on the growth rate measured from VIPERS data using our method comes from the variance of the value of $\beta$ recovered from individual VIPERS-like mocks multiplied by an additional factor, $m_2$,
\begin{equation}
\sigma_\beta^{\rm data} = \sqrt{m_2}\sigma_\beta^{\rm mocks},
\end{equation}
where
\begin{equation}
m_2 = \frac{m_1}{1-\mathcal{D}}.
\end{equation}
This additional factor accounts for the fact the mocks used to test the covariance matrix were also used to construct it. The VIPERS data is completely independent of the covariance matrix and thus is biased in a different way to the mocks.

\subsection{Measuring the bias}
\label{sec:bias}

In order to recover the growth rate corresponding to the input cosmology from the mocks, we must first estimate the effective linear bias of mock galaxies used to measure the void-galaxy cross-correlation function. Since we know the real space positions of galaxies in our mock catalogues we can measure the bias by taking the ratio of the real space correlation function of galaxies, $\xi_g(r)$, to the dark matter correlation function, $\xi_{dm}(r)$, 
\begin{equation}
b^2 = \frac{\xi_g(r)}{\xi_{dm}(r)}
\end{equation}
Here $\xi_{dm}(r)$ is the usual dark matter two-point autocorrelation function,
\begin{equation}
\xi(r) = \frac{1}{(2\pi)^3}\int P_{\rm dm}(k) \frac{\sin(kr)}{kr}4\pi k^2 {\rm d}k.
\end{equation}
This can be calculated by performing a Fourier transform of the theoretical dark matter power spectrum, $P_{\rm dm}(k)$, generated using {\sc camb} \citep{Lewis:2002ah}. The power spectrum has the same cosmological parameters as the mocks and is calculated at the median redshift of void-galaxy pairs, which is $z = 0.727$ (see Section \ref{sec:measurez}). The non-linear component of the matter power spectrum is estimated using {\sc halofit} \citep{2012ApJ...761..152T}.
\begin{figure}
\begin{center}
\includegraphics[width=0.5\textwidth]{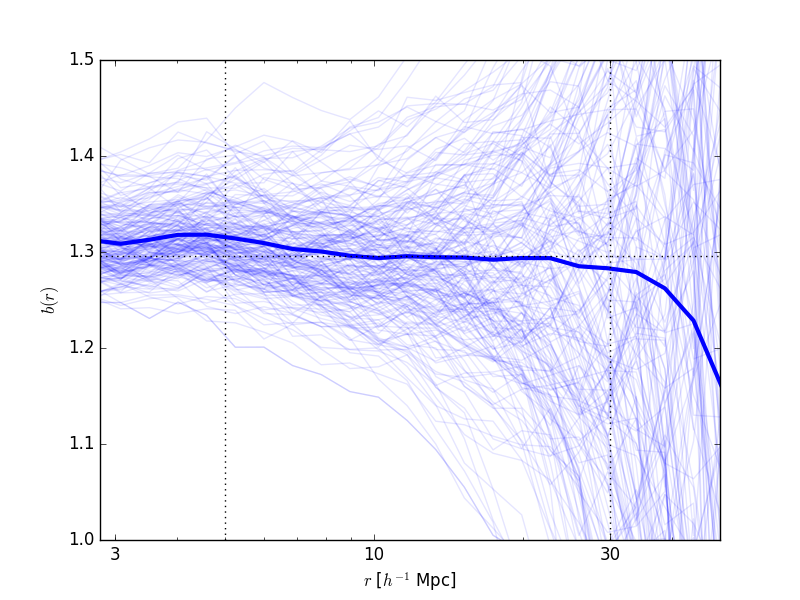}
\caption{Bias of mock galaxy catalogues. The faint blue lines represent the measured bias of individual mocks, the thick blue line is the mean of the mocks. Our quoted value for the mean bias (dotted horizontal line) is the mean value between $5.0 < r < 30 \mpcoh$ which is the scale over which the bias shows the least scale dependence (dotted vertical lines). The downturn at large scales is caused by the integral constraint.}
\label{fig:cf_bias}
\end{center}
\end{figure}
Having access to the real space positions of the mock galaxies, we measured the real space correlation function in the mocks using the Landy-Szalay estimator \citep{1993ApJ...412...64L},
\begin{equation}
\xi_g(r) = \frac{DD(r) - 2DR(r) + RR(r)}{RR(r)},
\end{equation}
where $DD(r)$ is the number of galaxy-galaxy pairs in a given bin of comoving separation, $r$, $DR(r)$ is the number of galaxy-random pairs, and $RR(r)$ is the number of random-random pairs. The bias measured in the mock catalogues is plotted in Figure \ref{fig:cf_bias}. The bias has some scale dependance so we take an average value. The mean bias in the mocks over the scales $5.0 \leq r_p \leq 30.0$ and its error are $b =  1.29 \pm 0.02$. The mean error for one mock is $0.05$.

\subsection{Testing on the toy model}
\label{sec:tst_tm}
We first tested the method on the toy model for the density profile presented in Section \ref{sec:model}. We wish to ensure that our method of deprojecting the cross-correlation function to estimate the void density profile does not introduce a bias on the measured growth rate. By applying our RSD model we generated an anisotropic cross-correlation function from the toy model, with a known value of $\beta=0.64$ and fixing $\sigma_v=13.4$. We then treated this in the same way we would treat data. We calculated the toy model in $25\times25$ bins and then deprojected it to obtain an estimate of the input model density profile. % Figure \ref{fig:1dmodel} shows that the deprojected cross-correlation (red dashed line) closely matches the toy model (blue solid line). 
We found that reducing the number of bins from which the deprojected cross-correlation is measured can introduce an offset. We then ran an MCMC chain on the toy model. %This was then reanisotropised for different values of $\beta$ and the likelihood of each value calculated.% Figure \ref{fig:mcmc_toy_model} shows the likelihood as a function of $\beta$. 
The true value of $\beta$ was well recovered, with minimal bias being introduced by the method.

\subsection{Recovering the input cosmology from the mocks}
\label{sec:mock_tests}
In order to demonstrate that the model presented in Section \ref{sec:rsd} is a sufficient description of the anisotropic void-galaxy cross-correlation function, we must show that we are able to extract the correct growth rate of structure from the mock galaxy catalogues described in Section \ref{sec:mocks}. 
This is a test both of our method and our RSD model.

\begin{figure*}
\begin{center}
\includegraphics[width=\textwidth]{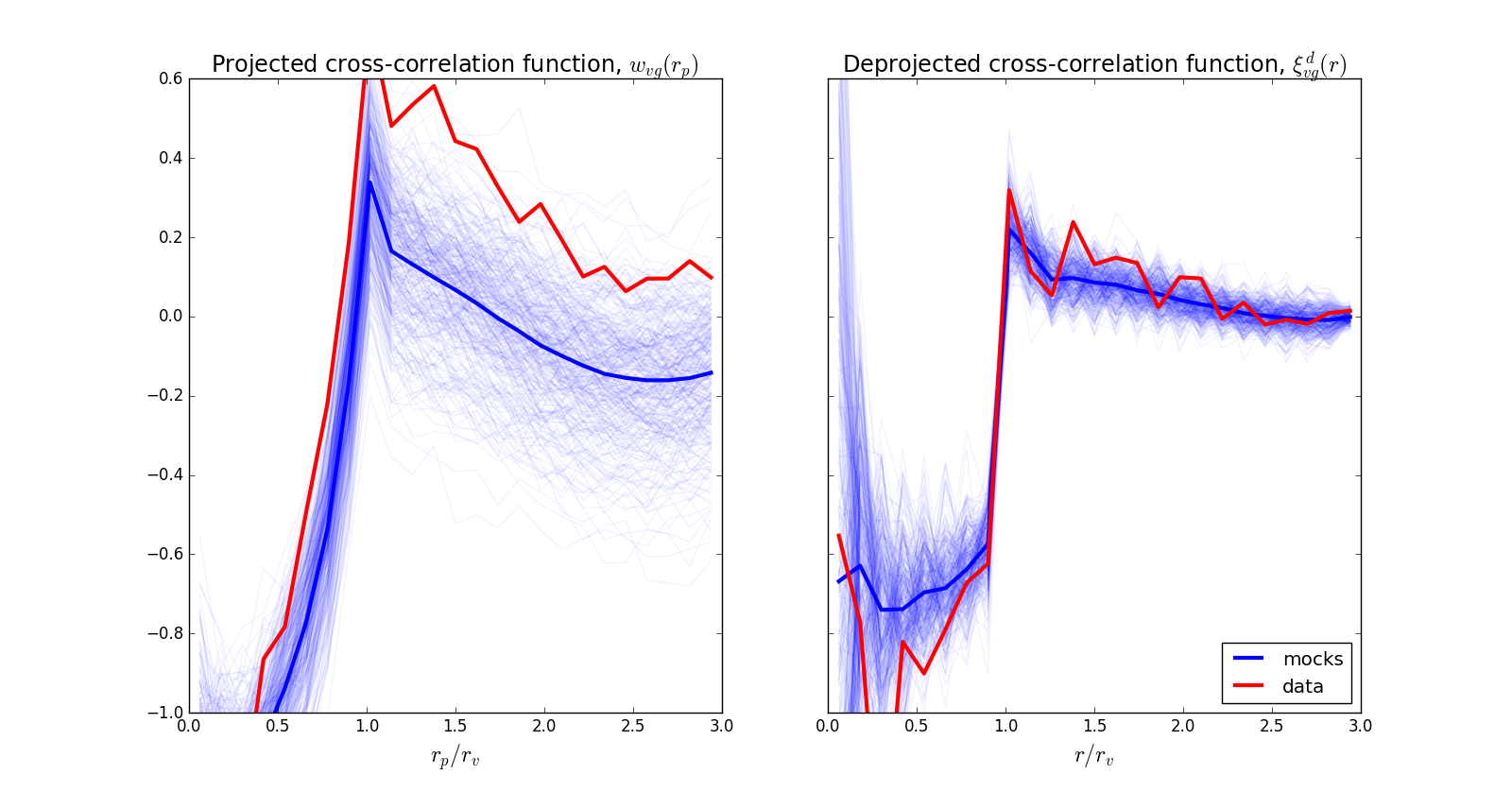}
\caption{Projected (left hand panel) and deprojected (right hand panel) void-galaxy cross-correlation functions for mock catalogues (blue) and the VIPERS data (red). Here the minimum void radius used is $8\mpcoh$. The mock catalogues were not constructed with a mind to accurately reproducing void properties, therefore the fact that there are some inconsistencies between mock and data void profiles is to be expected.}
\label{fig:profiles}
\end{center}
\end{figure*}

The projected cross-correlation functions and the deprojected cross-correlation functions of all 306 mocks are plotted in Figure \ref{fig:profiles} (left and right hand panels respectively). The thick blue line in each panel represents the mean value. ($w_{vg}(r_p)$ and $\xi^{d}_{vg}$ for the data are also plotted; these will be discussed in the next section.)

\begin{figure*}
\begin{center}
\includegraphics[width=\textwidth]{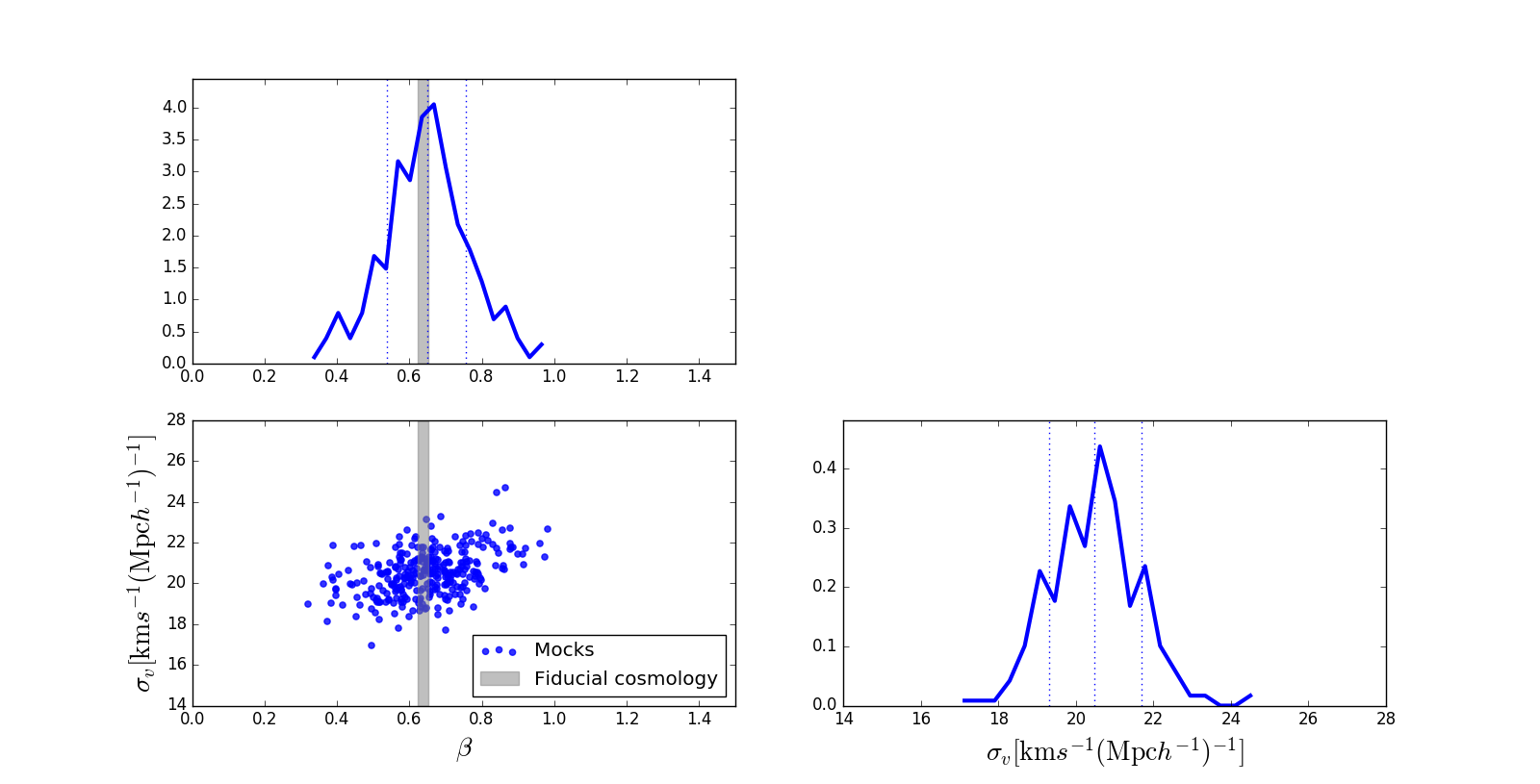}
\caption{Distribution of recovered values of $\beta$ and $\sigma_v$ from mock catalogues. Each blue point in the bottom left panel gives the best fitting values of $\beta$ and $\sigma_v$ for the combination of two VIPERS-like mock fields. The histogram in the top panel shows the PDF of the recovered values of $\beta$ and the bottom right panel gives the PDF of the recovered values of $\sigma_v$. The grey band is the expected value of $\beta$ given the fiducial cosmology and the uncertainty on the bias.}
\label{fig:mockdist}
\end{center}
\end{figure*}

We then ran {\sc emcee}, an implementation of the affine-invariant ensemble sampler for Markov chain Monte Carlo algorithm \citep{2013PASP..125..306F}, to estimate the best fitting values of $\beta$ and $\sigma_v$ in each of the 306 mock realisations. 
An accurate estimate of the uncertainty on the growth rate measured from VIPERS using our method comes from the distribution of the value of $\beta$ recovered from individual VIPERS-like mocks. This also allows us to place a non-Gaussian error bar on our result. Figure \ref{fig:mockdist} shows the scatter of recovered values of $\beta$ and $\sigma_v$ for the mocks. The top panel shows histogram of recovered values of  $\beta$, the grey band shows the expected value of $\beta$ given the cosmology and bias of the mocks. The 16th and 84th percentiles are illustrated by the dotted blue lines. The true value of $\beta$ lies very close to the mean of those recovered from the mocks. The distribution of recovered values is not strongly non-Gaussian.

\section{Application to VIPERS data}
\label{sec:vipersgrowth}
In this section we describe the application of the method tested on our mock catalogues in Section \ref{sec:growthrate} to the final data release of VIPERS. Section \ref{sec:measurez} describes how we estimate the redshift at which our measurement of $\beta$ is made. 
Section \ref{sec:datagrowthrate} presents our measurements of $\beta$ in the data. Section \ref{sec:comparison} then describes how we convert our measurement of $\beta$ to a measurement of $f\sigma_8$ so that it can be compared to other measurements in the literature.
\subsection{Estimating the redshift of the measurement}
\label{sec:measurez}
\begin{figure}
\begin{center}
\includegraphics[width=0.5\textwidth]{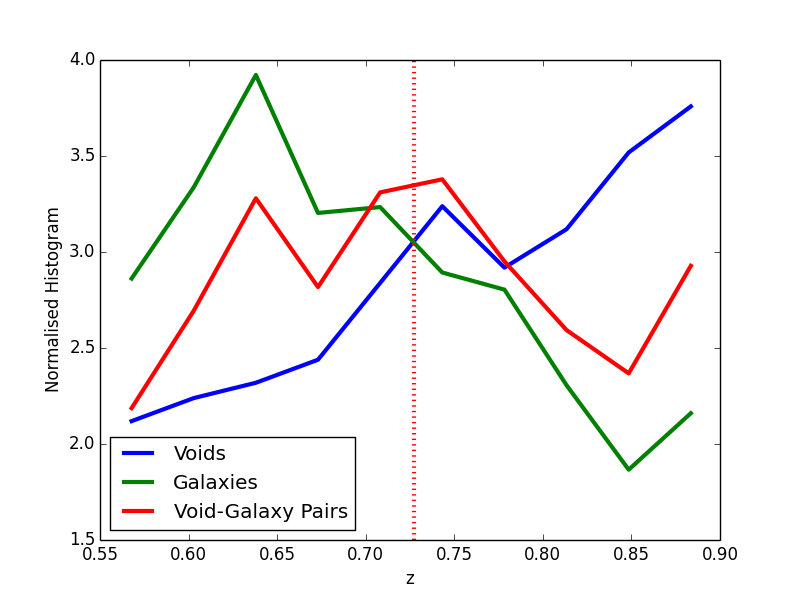}
\caption{Normalised number of objects as a function of redshift, $N(z)$, for voids (blue), galaxies (green), and void-galaxy pairs (red) in VIPERS. The mean redshift of void-galaxy pairs is $\bar{z}=0.727$ (dashed red line).}
\label{fig:selection_functions}
\end{center}
\end{figure}

It is important to note that our galaxy and void samples span a considerable distance in redshift space, $0.55 < z < 0.9$. The growth rate of structure is expected to evolve over this redshift range. The mean redshift at which we are measuring the growth rate will be some weighted combination of the radial selection functions of galaxies and voids, approximately the mean redshift of void galaxy pairs. Figure \ref{fig:selection_functions} shows the normalised number of objects as a function of redshift, $N(z)$, for our void catalogue (blue line) and for the full galaxy sample (green line). The $N(z)$ of voids rises with redshift, chiefly because there is more volume available at higher redshifts. The $N(z)$ of void galaxy pairs is then the product of these two histograms (red line). The red dashed line shows the mean redshift of pairs, $\bar{z}=0.727$, this is the redshift at which our measurement of the growth rate is made. 

\subsection{The effect of tracer luminosity on void properties}
\begin{figure}
\begin{center}
\includegraphics[width=0.5\textwidth]{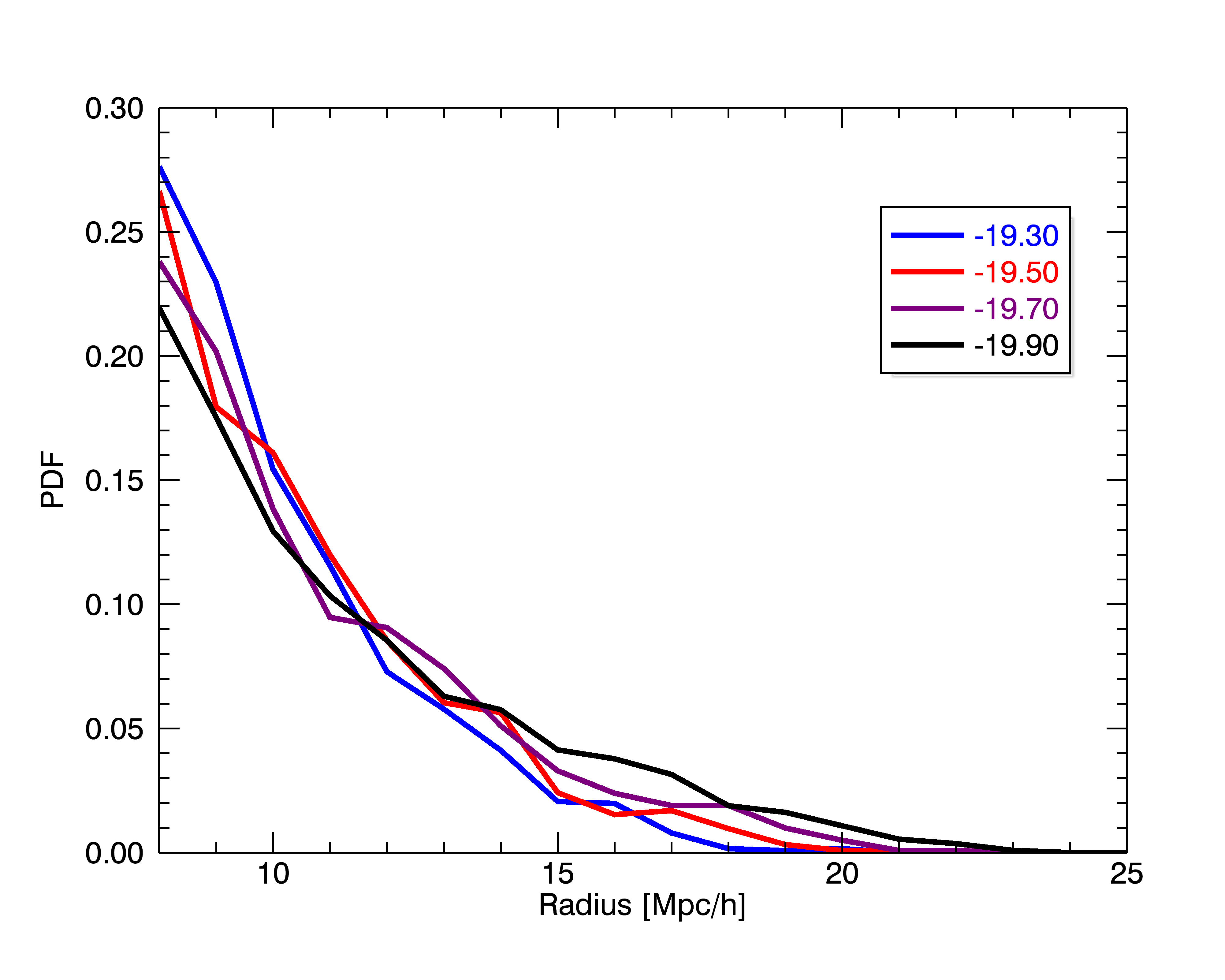}
\caption{Histograms showing the distribution of the size of void radii for voids found in different volume limited catalogues.}
\label{fig:magcuts_size}
\end{center}
\end{figure}

There are some minor differences between apparent and absolute magnitudes in different VIPERS data releases. We also know that the redshift evolution of absolute magnitudes in the mocks is not representative of the data. It is therefore useful to investigate what impact changing the magnitude limit, of the volume limited catalogue in which we search for voids, could have on our measurement of the void density profile. To do this we reran our void finder on volume limited catalogues with brighter magnitude cutoffs. Histograms showing the distribution of void radii in these samples are shown in Figure \ref{fig:magcuts_size}. As one might expect, more luminous tracers (and thus probably more biased tracers) define larger voids. This also means that fewer voids are found in these catalogues, and thus the signal to noise ratio of any statistics will be reduced.

\begin{figure*}
\begin{center}
\includegraphics[width=\textwidth]{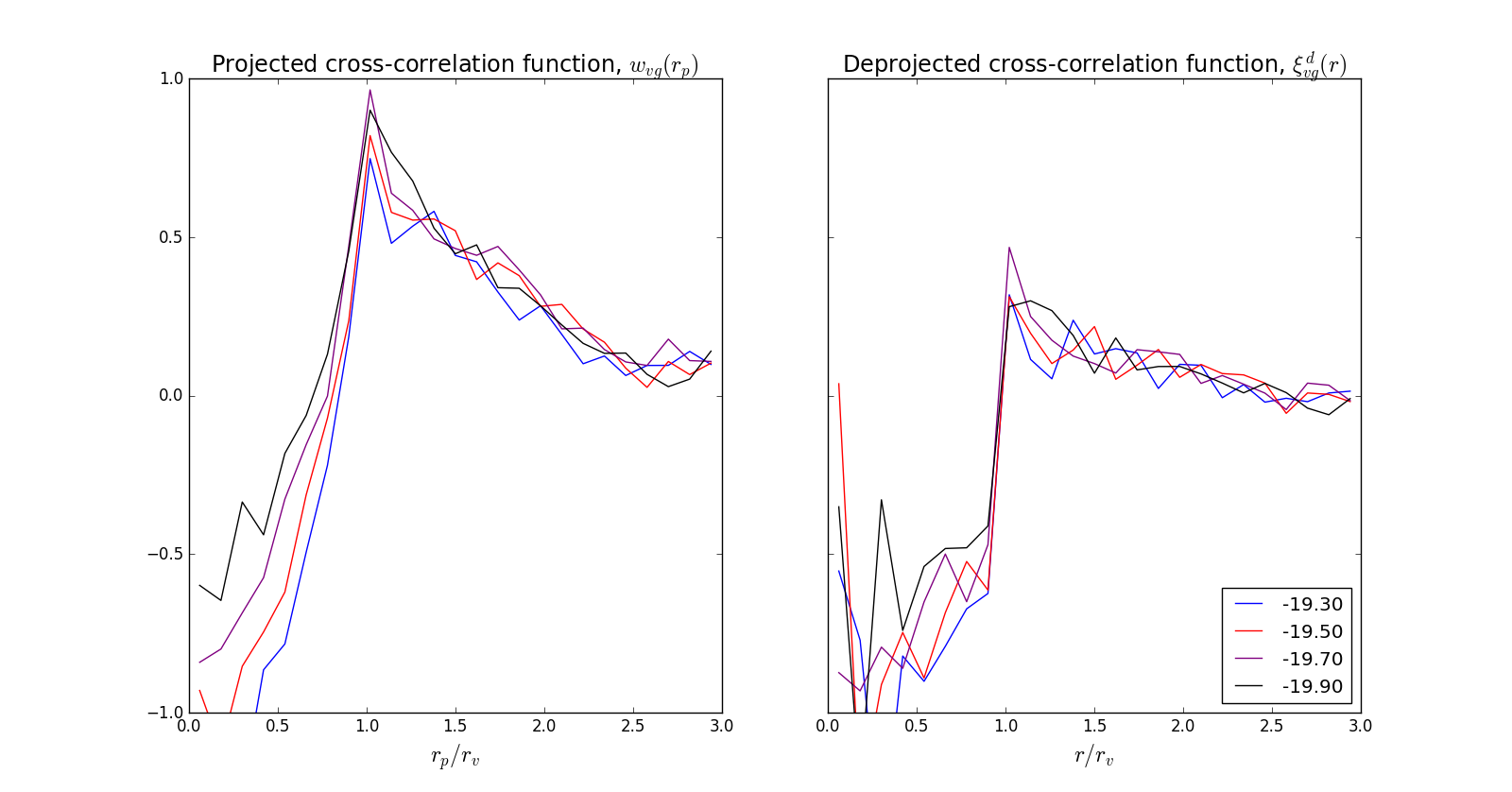}
\caption{Projected cross-correlation functions (left hand panel) and deprojected density profiles (right hand panel) for voids in VIPERS, found in volume limited catalogues with different magnitude cuts. When a brighter magnitude cut is used to define the volume limited catalogue the voids found are less empty and thus the interior void profile changes.}
\label{fig:magcuts_ccor}
\end{center}
\end{figure*}

We then measured the cross-correlation between voids found in these brighter samples and the complete galaxy population (as described in Section \ref{sec:measurements}). The corresponding projected cross-correlation functions and deprojected density profiles are plotted in Figure \ref{fig:magcuts_ccor} (left and right hand panels respectively). The larger voids defined by the brighter tracer populations have less underdense interiors. Other than that there is no clearly discernible trend. It is perhaps surprising that the brightness of the magnitude cut does not have a clear effect on the deprojected density profile.

\subsection{Estimating the growth rate}
\label{sec:datagrowthrate}

\begin{figure}
\begin{center}
\includegraphics[width=0.5\textwidth]{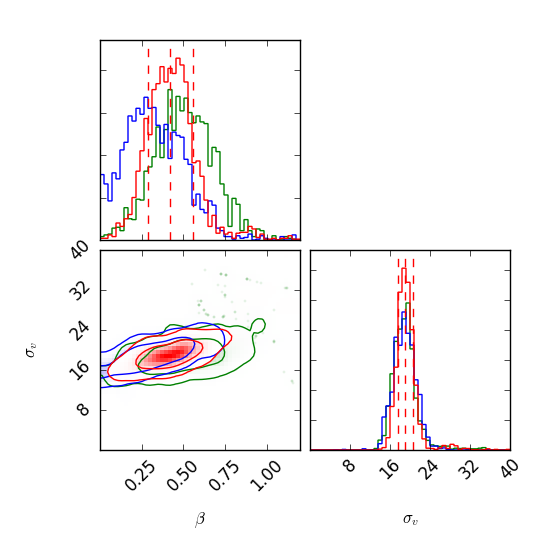}
\caption{MCMC contours for two parameter RSD model fit to VIPERS. The green contours indicate the fit to W4, the blue to W1, and the red contours are from the combination of the two fields. There is no significant tension between the two fields.}
\label{fig:mcmc_data}
\end{center}
\end{figure}

Using the method described in Section \ref{sec:growthrate} we fitted our model for the void-galaxy cross-correlation to the two VIPERS fields individually and to the combination of the two fields. 
Table \ref{tab:results} shows the best fitting values for $\beta$ and $\sigma_v$ and their associated errors as estimated using an MCMC chain.  

\begin{table}[]
\centering
\caption{Best fitting parameters to the data, as estimated using an MCMC chain. Errors on the estimated values are those from the MCMC. For the full VIPERS we also add errors estimated from the scatter of the mocks.}
\label{tab:results}
\begin{tabular}{lll}
\hline
      & $\beta$                   & $\sigma_v$  $[{\rm km \, s}^{-1} \,(\mpcoh)^{-1}]$        \\ \hline
W1    & $0.315^{+0.202^{\phantom{A}}}_{-0.162_{\phantom{A}}}$ & $18.9^{+2.2^{\phantom{A}}}_{-2.1_{\phantom{A}}}$ \\
W4    & $0.505^{+0.181^{\phantom{A}}}_{-0.175_{\phantom{A}}}$ & $18.8^{+2.0^{\phantom{A}}}_{-2.0_{\phantom{A}}}$ \\
VIPERS & $0.423^{+0.134\,(+0.104)^{\phantom{A}}}_{-0.135\,(-0.108)_{\phantom{A}}}$ & $19.1^{+1.6^{\phantom{A}}}_{-1.5_{\phantom{A}}}$ \\
\hline
\end{tabular}
\end{table}

The uncertainties quoted in the above table come from the likelihood and they misestimate the true uncertainty in the measurement. The analysis of the mock catalogues presented in Section \ref{sec:mock_tests} suggests that the error bar on the total measurement should be slightly smaller (though comparable). $\beta_{\rm VIPERS} = 0.423^{+0.104}_{-0.108}$. There is no significant inconsistency between the results from the two VIPERS fields.

Figure \ref{fig:mcmc_data} shows the contours from the MCMC analysis. It would suggest that there is a slight degeneracy between the two parameters. This degeneracy is also suggested by the scatter of best fitting values in the mock catalogues (Figure \ref{fig:mockdist}). However, given that the degeneracy is not steep, fixing $\sigma_v$ would only have a marginal effect on the error on the measured growth rate. Nevertheless, additional prior information about the velocity dispersion of galaxies around voids would aid in further constraining the growth rate.

\subsection{Comparison with other estimates of the growth rate}
\label{sec:comparison}
Conventionally, measurements of the growth rate of structure are quoted in terms of $f\sigma_8$, which is related to our measurement of $\beta$ by,
\begin{equation}
\label{eqn:fsig8}
f\sigma_8 = \beta \sigma_8^{\rm galaxies}.
\end{equation}
The values of \smash{$\sigma_8$} on the left and right hand side of the above equation are the {\it linear} values of $\sigma_8$ for dark matter and galaxies respectively.
Thus in order to compare our measurement of $\beta$ in VIPERS with other growth rate measurements we must also measure the value of $\sigma_8$ of galaxies in the data. The real space, {\it nonlinear}, $\sigma_8$ of galaxies can be estimated from the projected galaxy autocorrelation function \citep{2005ApJ...621...22Z, 2003ApJ...586..718E},
\begin{equation}
\label{eqn:sig8}
\sigma_R^2 = \frac{1}{R^3}\int_0^\infty r_p \, w_p(r_p) \, g(r_p/R) \, {\rm d}r_p,
\end{equation}
where $R=8 \mpcoh$ and
\begin{equation}
g(x) = \begin{cases} \frac{1}{2\pi}[3\pi - 9x + x^3] &\mbox{if } x \leq 2 \\ 
\frac{1}{2\pi}\bigg[\frac{-x^4 + 11x^2 -28}{\sqrt{x^2-4}} + x^3 -9x + 6\sin^{-1}\bigg(\frac{2}{x}\bigg)\bigg] & \mbox{if } x > 2. \end{cases} 
\end{equation}
The projected correlation function is defined as 
\begin{equation}\label{eqn:wp}
w_p(r_p) = 2 \int_{r_p}^\infty \frac{r \, \xi(r)}{\sqrt{r^2 - r_p^2}} \, {\rm d}r = 2 \int_0^\infty \xi \bigg(\sqrt{r_p^2 + r_\pi^2}\bigg) \, {\rm d}r_\pi,
\end{equation}
where $r$ is the apparent comoving separation of galaxy pairs, $r_\pi$ is the line-of-sight separation, and $r_p$ is their projected separation perpendicular to the line of sight. 
We measure $w_p(r_p)$ by using the Landy-Szalay estimator to measure $\xi(r_p, r_\pi)$ of galaxies and integrate it using Equation (\ref{eqn:wp}). In practice, the limits of the integral in \ref{eqn:wp} are finite and determined by observational constraints. On scales $r<1 \mpcoh$ the galaxy autocorrelation function is dominated by systematic effects, namely the TSR and SSR \citep[see][]{delatorreetal13}. We cannot measure scales $r_\pi \gg 100 \mpcoh$ due to the finite size of the survey. The limits of the integral are thus taken to be, $1 \mpcoh < r_\pi < 120 \mpcoh$. This result is then integrated using Equation (\ref{eqn:sig8}) to obtain an estimate of $\sigma_8^{\rm galaxies}$. 

The {\it linear} value of \smash{$\sigma_8^{\rm galaxies}$} can then be estimated by multiplying by the factor \smash{$\sigma_8^{\rm linear}/\sigma_8^{nonlinear}$}, the ratio of the linear and non-linear values for the \smash{$\sigma_8$} of dark matter, calculated from a {\sc CAMB} power spectrum, respectively without and with a {\sc halofit} model for the non-linear part. The ratio \smash{$\sigma_8^{\rm linear}/\sigma_8^{nonlinear}$} is fairly model independent, so the use of a fiducial power spectrum should not affect our result. However, using the ratio computed for dark matter to estimate the same ratio for galaxies implicitly assumes linear biasing. 

\begin{figure}
\begin{center}
\includegraphics[width=0.5\textwidth]{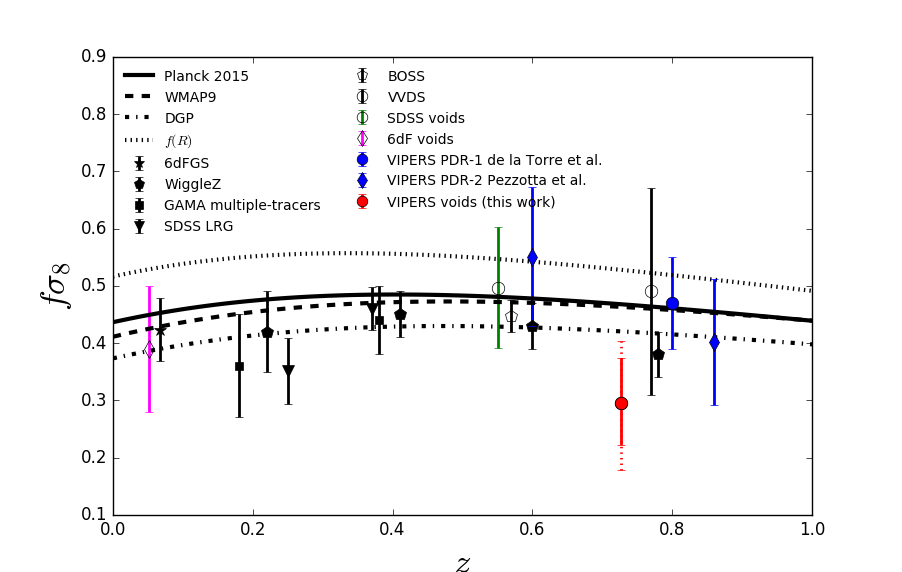}
\caption{Comparison to other estimates of the growth rate \citep{2012MNRAS.423.3430B, 2011MNRAS.415.2876B, 2013MNRAS.436.3089B, 2012MNRAS.420.2102S, 2014MNRAS.439.3504S, 2008Natur.451..541G}. Of particular interest are the measurement using conventional galaxy clustering techniques on VIPERS PDR-1 \citep[blue filled circle: ][]{ delatorreetal13}; the measurement using voids in SDSS \citep[green open circle: ][]{2016arXiv160201784H}; and the measurement using voids in 6dF \citep[magenta diamond: ][]{2016arXiv160603092A}.}
\label{fig:growth_plot}
\end{center}
\end{figure}

We measured a mean value of \smash{$\sigma_8^{\rm galaxies} = 0.735 \pm 0.043$} in our mock catalogues (consistent with the estimate of the bias presented in Section \ref{sec:bias}). The value recovered from the data is \smash{$\sigma_8^{\rm galaxies} = 0.700$}. Our estimate of $f\sigma_8$ is then: \smash{$f\sigma_8 = 0.296^{+0.075}_{-0.078}$}. Figure \ref{fig:growth_plot} shows this value compared to other measurements. 

\section{Discussion and Conclusion}
\label{sec:conclusion}

With the final data set of VIPERS we produced an updated void catalogue. We measured the anisotropic cross-correlation between the centres of voids in this catalogue and the full VIPERS galaxy sample. By deprojecting the anisotropic cross-correlation we were able to estimate the undistorted density profile. We demonstrated, first using a toy model and then using mock galaxy catalogues, that by fitting a model which includes linear redshift space distortions to the cross-correlation we can recover an estimate of the linear growth rate parameter $\beta$. Applying this to the combined data set of the two VIPERS fields we obtained a measurement of \smash{$\beta_{\rm VIPERS} = 0.423^{+0.104}_{-0.108}$}. We can convert this to a value for the linear growth rate of \smash{$f\sigma_8 = 0.296^{+0.075}_{-0.078}$}.

There is no significant tension between our measurement and that obtained from a conventional analysis of the VIPERS data, although our measurement appears to be slightly lower. Our measurement is commensurate with other published results using more conventional methods. 

The dominant source of uncertainty is cosmic variance. The usefulness of the void-galaxy cross-correlation function from VIPERS for constraining cosmology is limited by the size and geometry of the survey. Since our mock catalogues have a VIPERS-like geometry, we cannot investigate possible  constraints from a larger contiguous region and are restricted to studying scenarios with VIPERS-like fields. It is likely that a larger contiguous survey would provide much tighter constraints.

Our algorithm rejects spheres when less than 80\% of the volume falls within the survey. One of the results of this is that close to the borders of the survey voids can become fragmented, with large spheres being replaced by many smaller ones. Border effects are not unique to our algorithm: ZOBOV based void finders also have problems describing voids close to survey boundaries \citep{2008MNRAS.386.2101N}. A popular approach to dealing with this problem is to exclude voids which lie close to the borders from the analysis. However, the geometry of VIPERS makes it particularly susceptible to border effects. In Figure \ref{fig:cc_all}, the signal from W4 appears noisier, by eye, than the signal from W1. It is worth pointing out that, being smaller, W4 will be more affected by border effects than W1. Almost all voids intersect with at least one survey boundary, so excluding voids which intersect with borders from the analysis would be unfeasible. 

Our model for the redshift space distortions around voids, outlined in Section \ref{sec:rsd}, assumes that the centres of empty spheres correspond to maxima in the gravitational potential field, i.e. points from which galaxies are outflowing. Although our results clearly indicate a positive detection of outflows from voids, it may well not be the case that the centres of our spheres correspond to the centres of these outflows. Any random offset is likely to dilute the redshift space distortion signal and add to the uncertainty in the estimate of $\beta$ - but this will be allowed for in mocks, and we see no such effect.

If it is the case that the properties of galaxies in the void interiors are significantly different to those outside, then they will be biased with respect to the dark matter distribution in different ways. In this paper we have assumed that the galaxy bias is strictly linear and scale independent. A more thorough model for the velocity field should consider scale dependent bias around voids \citep{2014MNRAS.441..646N}.

To date there are two other works to have attempted measuring $\beta$ from the void-galaxy cross-correlation in data: they are \citet{2016arXiv160603092A} and \citet{2016arXiv160201784H} [green and magenta points of Figure \ref{fig:growth_plot}]. These results were released whilst our analysis was being carried out.

There are several key differences between the work of \citet{2016arXiv160201784H} and ours. In terms of methodology, rather than directly deprojecting the void density profile they assume a certain functional form for it and then marginalise over the parameters of their model. The  Sloan Digital Sky Survey covers a much larger volume than VIPERS, thus \citet{2016arXiv160201784H} have many more galaxy-void pairs from which to measure the cross-correlation. They also probe different scales to us. Their voids range in size from $24\mpcoh$ to $64\mpcoh$. The largest void in our analysis has a radius of $20.8\mpcoh$, smaller than their smallest void, whilst their largest void bin is comparable to the width of VIPERS. This could have an impact on the accuracy of our redshift space distortion model, since it is understood that velocity fields of smaller voids are less linear than those of larger ones. It can therefore be expected that a linear description of the velocity field around voids is a less good description for a survey such as VIPERS than for SDSS. However, any changes to the recovered growth rate from improved modelling are likely to remain within the current error bar.

\citet{2016arXiv160603092A} looked at the void-galaxy cross correlation in the 6dF survey. They take an undistorted $\xi_{vg}$ calibrated on dark matter simulations and fit it to the anisotropic cross-correlation. Their algorithm is able to select voids of a certain size, $\sim20\mpcoh$ fitting a particular profile. Some of their voids overlap, while ours are defined not to. They exclude some bins on small scales to mask out nonlinearities. The number of spectra measured in the 6dF survey is of the same order of magnitude as that measured by VIPERS.

\citet{2016arXiv160305184C} present a method for measuring the linear growth rate $\beta$ using the multipoles of the void-galaxy cross-correlation function. They then apply this method to simulations and demonstrate that given a volume of $3{\rm Gpc}^3h^{-3}$ they can recover $\beta$ to within $10\%$. Their methodology has some similarities to ours. Firstly they define their voids using underdense spheres, as do we. Secondly their approach does not require a model for the void density profile, since they are able to derive this from the multipoles. There are some differences in their redshift space distortion modelling, for most of their analysis they ignore the velocity dispersion, $\sigma_v$, and correlations close to the void centres. However, when they include $\sigma_v$ and the void interiors they are able to reduce the uncertainty of $\beta$.

The precision of our measurement is consistent with the precision of \citet{2016arXiv160603092A} and is better than that of \citet{2016arXiv160201784H}, given the difference in survey volume. Although VIPERS may not provide the most accurate measurement of the growth rate of structure in low density environments, it provides a measurement at higher redshift than other current observations. Thus our results limit any gross deviations from Einstein gravity at high redshift.

In a parallel paper of this series the growth rate of structure has been measured using a more conventional technique. Pezzotta et al. (2016) measured the growth rate by modelling the multipoles of the anisotropic autocorrelation in configuration space. They found $f\sigma_8 = 0.551\pm0.121$ and $0.401\pm0.110$ at $z = 0.6$
and $0.86$ respectively (blue diamonds Figure \ref{fig:growth_plot}). %de la Torre et al. (2016) combined RSD measurements from VIPERS with lensing data from the CFHTLenS survey over the same area of sky to constrain the growth rate of structure. They found $f\sigma_8 = 0.37 \pm 0.06$ and $f\sigma_8 = 0.47 \pm 0.06$ at $z=0.61$ and $0.85$ respectively (blue triangles Figure \ref{fig:growth_plot}). 
Our estimate for  is lower than those obtained from VIPERS in %de la Torre et al. (2016) and 
Pezzotta et al. (2016).  Estimating the growth rate from the void-galaxy cross-correlation
 function is clearly still in its infancy, with potential systematic
 errors not yet fully understood. Nevertheless, accounting for the different
 effective redshifts of the measurements, the different VIPERS values for
 the growth rate are consistent at the 1-sigma level.  

\section*{Acknowledgements}

We would like to thank Paul M. Sutter, Nico Hamaus, and Dante Paz for interesting and thought provoking discussions.

AJH, BRG, and LG acknowledge the
support of the European Research Council through
the Darklight ERC Advanced Research Grant (291521).

We acknowledge the crucial contribution of the ESO staff for the management of service observations. In particular, we are deeply grateful to M. Hilker for his constant help and support of this program. Italian participation to VIPERS has been funded by INAF through PRIN 2008 and 2010 programs. OLF acknowledges support of the European Research Council through the EARLY ERC Advanced Research Grant (\# 268107). AP, KM, and JK have been supported by the National Science Centre (grants UMO-2012/07/B/ST9/04425 and UMO-2013/09/D/ST9/04030), the Polish-Swiss Astro Project (co-financed by a grant from Switzerland, through the Swiss Contribution to the enlarged European Union). KM was supported by the Strategic Young Researcher Overseas Visits Program for Accelerating Brain Circulation No. R2405. GDL acknowledges financial support from the European Research Council under the European Community's Seventh Framework Programme (FP7/2007-2013)/ERC grant agreement n. 202781. WJP and RT acknowledge financial support from the European Research Council under the European Community's Seventh Framework Programme (FP7/2007-2013)/ERC grant agreement n. 202686. WJP is also grateful for support from the UK Science and Technology Facilities Council through the grant ST/I001204/1. EB, FM and LM acknowledge the support from grants ASI-INAF I/023/12/0 and PRIN MIUR 2010-2011. LM also acknowledges financial support from PRIN INAF 2012. YM acknowledges support from CNRS/INSU (Institut National des Sciences de l'Univers) and the Programme National Galaxies et Cosmologie (PNCG). CM is grateful for support from specific project funding of the {\it Institut Universitaire de France} and the LABEX OCEVU. Research conducted within the scope of the HECOLS International Associated Laboratory, supported in part by the Polish NCN grant DEC-2013/08/M/ST9/00664.

\bibliographystyle{aa}
\bibliography{growth_paper}

%\begin{thebibliography}{}
%
%\bibitem[Ade \etal\ 2014]{planck}Ade, P. A. R., \& The Planck Collaboration, \ 2014, \textit{A\&A}, 571, A16
%
%\bibitem[Aragon-Calvo \& Szalay 2013]{}Aragon-Calvo, M. A., Szalay, A. S., \ 2013, \textit{MNRAS}, 428, 4
%
%\bibitem[Dubinski \etal\ 1993]{dubinskietal}{Dubinski}, J., {da Costa}, L.~N., {Goldwirth}, D.~S., 
%	{Lecar}, M., \& {Piran}, T. \ 1993, \textit{ApJ}, 410
%
%\bibitem[Fimelli \etal\ 2014]{ltbvoid}{Finelli}, F., {Garcia-Bellido}, J., {Kovacs}, A., 
%	{Paci}, F., {Szapudi}, I., \ 2014, ArXiv e-prints:1405.1555
%
%
%
%
%\bibitem[Kaiser 1987]{kaiser}{Kaiser, N.} 1987, \textit{MNRAS}, 227
%

%
%
%\bibitem[Padilla \etal \ 2005]{padillaetal}Padilla, N. D., Ceccarelli, L., \& Lambas, D. G. \ 2005, \textit{MNRAS}, 363, 977
%
%\bibitem[Paranjape \etal\ 2012]{}Paranjape, A., Lam, T. Y., Sheth, R. K., \ 2012, \textit{MNRAS}, 420, 1648
%
%\bibitem[Patil \etal\ 2010]{pymc}Patil, A., Huard, D., Fonnesbeck, C. J., \ 2010, Journal of Statistical Software, 35, 4
%
%
%\bibitem[Prada \etal\ 2012]{multidark}{Prada}, F., {Klypin}, A.~A., {Cuesta}, A.~J., {Betancort-Rijo}, J.~E., \& {Primack}, J., \ 2012 ,\textit{MNRAS}, 423
%
%\bibitem[Regos \& Geller 1991]{regosandgeller}Regos, E. \& Geller, M. J. \ 1991, \textit{ApJ}, 377, 14
%
%
%\bibitem[Sutter \etal\ 2015]{vide}
%{Sutter}, P.~M., {Lavaux}, G., {Hamaus}, N., {Pisani}, A., 
%	{Wandelt}, B.~D., {Warren}, M., {Villaescusa-Navarro}, F., 
%	{Zivick}, P., {Mao}, Q. \& {Thompson}, B.~B.}, \ 2015, \textit{Astronomy and Computing}, 9
%
%\end{thebibliography}

\end{document}